\documentclass[12pt]{article}
\usepackage{graphicx}
\usepackage{authblk}
\usepackage[margin=1in]{geometry}
\usepackage{amsmath, amssymb}
\usepackage{hyperref}

\title{Quantum Bayesian Inference with State Vectors for Intrusion Detection}

\author[1]{Nayema Mridha}
\author[5]{Garrv Sipani}
\author[6]{Eva R. Gaarder}
\author[3]{Shah Haque}
\author[3]{Radhika Kuttala}
\author[4]{Binay Prakash Akhouri}
\author[3]{Mohamad Mahmoud Al Zein}
\author[1,2,3]{Eric Howard}

\affil[1]{Department of Computer Science, Macquarie University, Sydney, NSW 2109, Australia}
\affil[2]{School of Mathematical and Physical Sciences, Macquarie University, Sydney, NSW 2109, Australia}
\affil[3]{Southern Cross Institute, School of Computer Science, Sydney, Australia}
\affil[4]{Department of Physics, Suraj Singh Memorial College, Ranchi University, Ranchi, Jharkhand, India}
\affil[5]{Department of Cyber Security, Johns Hopkins University, Baltimore, Maryland 21210, USA}
\affil[6] {Department of Engineering, University of Technology Sydney, Sydney, NSW 2007, Australia}

\date{June 2025}

\begin{document}

\maketitle

\begin{abstract}
\begin{abstract}
We present a quantum Bayesian inference method for intrusion detection, using explicitly constructed quantum circuits and statevector simulation. Prior and conditional probabilities are encoded via unitary gates, and posterior distributions are extracted through symbolic post-selection. Applied to a scenario with network spikes, system vulnerabilities, and false alarms, the method yields joint, marginal, and conditional probabilities aligned with causal structure. Our results demonstrate the feasibility and interpretability of quantum-native inference for information security applications.
\end{abstract}

\end{abstract}

\section{Introduction}
Intrusion Detection Systems (IDS) are critical components in the modern cybersecurity infrastructure, tasked with monitoring and identifying potential threats such as Distributed Denial of Service (DoS) attacks, unauthorized access attempts, and false alarms triggered by benign anomalies. Traditional IDS approaches rely on either rule-based mechanisms or machine learning models that process observed network behavior to infer possible security breaches. These methods, while powerful, face increasing challenges as network traffic grows in complexity and as attackers employ more sophisticated evasion techniques. The inherent probabilistic nature of intrusion events calls for models capable of representing uncertainty and interdependent causal relationships more effectively.

Bayesian inference, a well-established framework in probabilistic reasoning, offers a structured approach for updating beliefs in light of new evidence. Within the context of intrusion detection, Bayesian models can represent the dependencies between observed network spikes, underlying system vulnerabilities, and the classification of alarms. However, classical Bayesian inference methods encounter computational bottlenecks when scaling to networks with many variables and dependencies, particularly in real-time or near-real-time environments. Quantum computing, with its ability to represent and manipulate large state spaces in superposition, provides an alternative paradigm for performing probabilistic inference. Recent advances in quantum machine learning continue to expand its algorithmic scope and application domains \cite{chowdhury2024, howard2021}.

In this study, we explore a quantum-native implementation of Bayesian inference using the \texttt{Statevector} representation in Qiskit. Our approach involves manually constructing the quantum circuits that encode the prior and conditional probability distributions. We then simulate the evolution of the quantum state, extract the resulting joint probability amplitudes, and compute posterior distributions by conditioning on observed evidence. This method allows for granular control over the inference process and provides greater flexibility in modeling custom security scenarios.

The specific use case addressed involves modeling the relationships among three key variables: a network traffic spike (X), the presence of a system vulnerability (Y), and the classification of the event as a false alarm (FA). These variables are encoded as qubits, and their interactions are modeled through a quantum circuit that reflects the causal structure of the Bayesian network. By analyzing the simulated statevector, we derive the likelihood of different events, such as the system being vulnerable given a spike, or the alarm being a false positive.

This approach not only demonstrates the feasibility of using statevector-based inference for cybersecurity applications but also opens avenues for more efficient probabilistic modeling as quantum hardware continues to evolve. The ability to compute conditional probabilities through quantum simulations introduces a novel toolkit for security analysts and researchers seeking to augment classical systems with quantum-enhanced reasoning. Moreover, our method illustrates the transparency and customizability of explicit inference, making it suitable for exploratory research and educational purposes in quantum information science.

The structure of this paper is as follows. Section 2 presents the scenario overview and variable mapping. Section 3 details the quantum Bayesian inference process, including circuit construction, simulation, and conditioning. Section 4 presents a cybersecurity use case, followed by visualizations and results in Section 5. The paper concludes in Section 6 with a discussion on the implications of quantum Bayesian inference for future IDS research and deployment.

The experiments in this study were conducted using the Qiskit software framework, a Python-based open-source platform developed by IBM for quantum computing. All code was executed in Google Colab, an interactive notebook environment that supports Python and provides free access to computational resources, including GPU acceleration when required. Colab’s compatibility with Qiskit enabled rapid development, visualization, and testing of quantum circuits directly in the cloud.

To perform quantum Bayesian inference, we utilized the `qiskit` library — particularly the `QuantumCircuit`, `QuantumRegister`, and `Statevector` classes. The inference pipeline was implemented without relying on prebuilt abstractions such as the `QBayesian` class, allowing full control over the quantum circuit design and probability interpretation.

Probability distributions were encoded into quantum states using Y-axis rotations (\texttt{ry} gates). Conditional dependencies between variables were modeled with controlled rotation gates (\texttt{cry}). Each quantum variable was mapped to a qubit, and the resulting quantum state was simulated using \texttt{Statevector.from\_instruction()}. From the statevector, we computed joint distributions, marginals, and conditionals through manual filtering and normalization.

Numerical computations, including trigonometric derivations of rotation angles, were handled using NumPy. Plotting and visualization were implemented with Matplotlib and Seaborn, providing heatmaps, bar charts, 3D plots, and CDFs to analyze simulation outcomes.

Our core workflow consisted of the following steps:
\begin{enumerate}
  \item Define the probabilistic model in terms of prior and conditional probabilities.
  \item Construct the quantum circuit that encodes these probabilities into qubit states.
  \item Simulate the quantum circuit using the \texttt{Statevector} method.
  \item Extract probability data for all possible qubit outcomes.
  \item Perform conditional filtering to implement inference.
  \item Visualize the results to interpret outcomes.
\end{enumerate}

\section{Mathematical Framework}
The mathematical foundation of our quantum Bayesian inference model is grounded in quantum statevector formalism, which represents classical probability distributions within quantum systems. Each binary variable in the probabilistic model (e.g., \(X\), \(Y\), \(FA\)) is encoded as a qubit, with the quantum state residing in a Hilbert space of dimension \(2^n\), where \(n\) is the number of qubits.

The global quantum state \(|\psi\rangle\) is a superposition of computational basis states:
\[ |
\psi\rangle = \sum_{i=0}^{2^n - 1} c_i |i\rangle \]
where \(c_i \in \mathbb{C}\) are complex amplitudes, and each \(|i\rangle\) denotes a basis state in the system. The squared modulus of each amplitude, \(|c_i|^2\), represents the probability of observing the corresponding basis state upon measurement. The normalization constraint is given by:
\[ \sum_{i=0}^{2^n - 1} |c_i|^2 = 1 \]

Marginal distributions are embedded via single-qubit rotations, specifically using \texttt{RY} gates. The rotation angle \(\theta\) is computed based on the target marginal probability \(p\), using the transformation:
\[ \theta = 2 \arcsin(\sqrt{p}) \]
This results in the qubit state:
\[ |q\rangle = \cos(\theta/2)|0\rangle + \sin(\theta/2)|1\rangle \]

To represent conditional dependencies, controlled rotation gates such as \texttt{CRY} are used. These gates apply conditional amplitudes based on the state of one or more control qubits. By carefully selecting the rotation angles, one can encode conditional probability tables (CPTs) consistent with classical Bayesian networks.

The overall joint probability distribution is constructed via a sequence of entangling and single-qubit gates. For example, if \(FA\) is conditionally dependent on \(X\) and \(Y\), we introduce multiple \texttt{CRY} gates acting on the FA qubit, with \(X\) and \(Y\) serving as controls. This results in a quantum state that captures all possible logical relationships specified by the CPT.

From the constructed statevector, marginal and conditional probabilities can be calculated. Marginals are obtained by summing squared amplitudes over basis states where the target variable satisfies a given condition. Conditional probabilities require filtering amplitudes according to the evidence and renormalizing:
\[ P(A \mid B) = \frac{\sum_{i: A,B} |c_i|^2}{\sum_{j: B} |c_j|^2} \]

Quantum inference in this context involves symbolic post-selection, where basis states satisfying the given evidence are isolated, and normalization over this subset gives the posterior distribution. This technique is theoretically equivalent to classical Bayes’ theorem, but leverages the compact representation of the full joint distribution in quantum amplitude space.

To evaluate inter-variable relationships, information-theoretic measures are defined. Shannon entropy quantifies uncertainty:
\[ H(Z) = -\sum_z P(z) \log_2 P(z) \]
while mutual information captures the dependency between two variables:
\[ I(X;Y) = \sum_{x,y} P(x,y) \log_2 \frac{P(x,y)}{P(x)P(y)} \]

Additionally, fidelity is introduced as a metric for quantifying similarity between two discrete probability distributions, particularly between an inferred distribution \(P\) and a reference distribution \(Q\):
\[ F(P,Q) = \left( \sum_i \sqrt{P_i Q_i} \right)^2 \]

Each quantum gate in the circuit represents a unitary transformation. When applying a gate to a single qubit within a multi-qubit system, identity operations are tensor-product extended to preserve the full Hilbert space. This ensures the modular and hierarchical assembly of the overall transformation. Interference among amplitudes in the superposition allows for subtle manipulation of correlations and anti-correlations. The principle of superposition provides a natural way to represent multiple outcomes simultaneously, enhancing the efficiency of inference over combinatorially large sample spaces.

The system evolution from the initial state \(|0\rangle^{\otimes n}\) to the final statevector is achieved through the application of a series of such unitary gates, which can be symbolically represented as:
\[ |
\psi\rangle = U_N \cdots U_2 U_1 |0\rangle^{\otimes n} \]
where each \(U_k\) corresponds to a logical operation encoding part of the model’s structure.

The theoretical apparatus is generalizable to higher-order Bayesian networks and can be adapted to any systems where probabilistic inference is required over large and structured variable sets.

\section{Scenario Overview}
We consider a cybersecurity monitoring system tasked with evaluating sudden spikes in network traffic. These spikes may indicate malicious activity, such as a Distributed Denial of Service (DoS) attack, or they may simply result from benign system behavior or pre-existing vulnerabilities. The challenge lies in determining the cause of the spike and the appropriate classification of the alert generated by the system. In particular, the system must distinguish between true threats and false alarms triggered due to inherent weaknesses in the infrastructure.

The scenario models three key variables:
\begin{itemize}
  \item \textbf{X (Network Spike)}: A binary variable indicating whether a significant spike in network traffic has been observed.
  \item \textbf{Y (System Vulnerability)}: A binary variable representing whether the system is known to be vulnerable to false positives.
  \item \textbf{FA (False Alarm)}: A binary variable denoting whether the alert raised by the IDS is classified as a false alarm.
\end{itemize}

These variables are encoded into a quantum circuit where each qubit corresponds to one of the three elements. The dependency structure reflects causal influence: the presence of a system vulnerability (Y) affects the likelihood of an alert being false (FA), and the occurrence of a network spike (X) can either be due to malicious intent or a side effect of a vulnerable system. Hence, the conditional probability of FA is determined by both X and Y, forming a Bayesian network structure.

The aim of the simulation is to compute conditional probabilities such as \( P(Y = 1 | X = 1) \) — the probability that the system is vulnerable given a spike — and \( P(FA = 1 | X = 1) \), the probability that the alarm is a false positive conditioned on the observed traffic anomaly. By building quantum circuits that encode the necessary joint distributions, we use statevector simulation to extract these inferences directly from the amplitudes of the quantum state.

This scenario represents a simplified yet practically relevant abstraction of how modern cybersecurity systems must reason under uncertainty. It captures the probabilistic nature of alerts, the influence of system weaknesses, and the need for dynamic classification of observed events. The model serves as a testbed for demonstrating how quantum computation, via statevector-based probabilistic modeling, can enhance decision-making in security domains.

\section{Quantum Bayesian Inference Process}
The process of quantum Bayesian inference using state vectors involves five key stages: probability encoding, quantum circuit construction, statevector simulation, probability extraction, and posterior computation. This section outlines the systematic methodology used to simulate a Bayesian network for intrusion detection on a quantum computer using Qiskit's \texttt{Statevector} formalism.

\subsection{Instantiating Quantum Bayesian Inference (QBI)}
Instantiating a quantum Bayesian inference model begins with encoding the prior and marginal probabilities of each binary variable into single-qubit rotations. Each binary variable, such as the presence of a network spike or system vulnerability, is mapped to a dedicated qubit initialized in the \(|0\rangle\) state.

For a given variable with prior probability \(p\), the target state is a superposition where the amplitude squared of \(|1\rangle\) matches \(p\). This is achieved through an \texttt{RY} rotation gate, where the rotation angle \(\theta\) is given by:
\[ \theta = 2 \arcsin(\sqrt{p}) \]
Applying \texttt{RY}(\(\theta\)) to a \(|0\rangle\) state yields the desired superposition:
\[ |q\rangle = \cos(\theta/2)|0\rangle + \sin(\theta/2)|1\rangle \]
This ensures \(|\langle 1|q\rangle|^2 = p\), aligning the quantum state's measurement probability with the classical marginal.

Multiple variables can be encoded simultaneously by assigning each to a separate qubit and applying corresponding \texttt{RY} rotations. The resulting quantum state is the tensor product of all individual qubits, forming a joint product state that represents the independent marginal distributions.

At this stage, no entanglement is present; each qubit evolves independently. This modular encoding is advantageous for clarity and modular circuit design, serving as a foundation for integrating conditional dependencies in the next phase. Additionally, the parameterization is adaptable — marginal probabilities can be dynamically updated by recalculating the rotation angles.

The design ensures that quantum amplitudes align with classical probabilities, forming a consistent bridge between classical Bayesian networks and quantum state preparation. This instantiation step is critical for maintaining interpretability and enabling subsequent logical conditioning steps to modify the probability landscape coherently.

The process also supports symbolic parameterization: one can use symbolic math libraries to define \(\theta\) as functions of parameter \(p\), facilitating circuit reconfiguration and gradient-based optimization for parameter learning, if desired.

By instantiating quantum Bayesian inference in this structured way, we ensure that the probabilistic semantics are preserved across the quantum pipeline, from initial qubit state preparation to inference via statevector analysis.

\subsection{Constructing Conditional Dependencies}
Constructing conditional dependencies in a quantum Bayesian network involves encoding classical conditional probability tables (CPTs) into quantum gates that operate on specific qubits. These conditional structures are critical for modeling relationships such as \(P(FA \mid X, Y)\), where the probability of a false alarm depends on both the observed network spike and system vulnerability.

To implement these conditional dependencies, we use controlled rotation gates—most commonly \texttt{CRY} gates—that rotate the target qubit (e.g., representing the variable FA) based on the state of one or more control qubits (e.g., X and Y). Each entry in the CPT corresponds to a specific configuration of the control qubits, with a predefined rotation angle that reflects the conditional probability.

In practice, when encoding \(P(FA \mid X=1, Y=0) = p\), the rotation angle \(\theta\) applied to the FA qubit is:
\[ \theta = 2 \arcsin(\sqrt{p}) \]
This ensures that the resulting quantum amplitude for the FA qubit being in state \(|1\rangle\) is \(\sqrt{p}\), and \(|c_i|^2 = p\) after measurement. Each combination of control qubit values maps to a corresponding quantum control gate that enforces the CPT relationship.

For multiple conditional variables, such as FA depending on both X and Y, multi-controlled gates (like \texttt{CCRY}) are decomposed into Toffoli-based networks to maintain universality and circuit depth optimization. These gates can be simulated directly using Qiskit’s decomposition or manually implemented using ancilla-assisted constructions.

Moreover, conditional dependencies introduce nontrivial entanglement across the qubit register, as the state of one variable influences the probability amplitude of another. This entanglement is essential for preserving the joint probabilistic structure across all variables in the network.

The design of these conditional gates must also respect topological and connectivity constraints of the target quantum architecture, although this consideration is more pertinent in hardware execution rather than simulation. Still, modularity in conditional gate construction supports circuit reusability and composability across different CPTs.

Overall, constructing conditional dependencies transforms a set of abstract probabilistic relationships into executable quantum instructions. It enables a flexible yet precise mapping from classical Bayesian networks into the quantum domain, preserving logical structure while exploiting quantum parallelism for inference tasks.

\subsection{Circuit Assembly}
With both marginal probabilities and conditional dependencies defined, the quantum circuit must be assembled in a manner that respects the causal ordering of the Bayesian network. Each gate is positioned such that parent variables are set before applying transformations to their dependent variables. This adheres to the topological ordering dictated by the network’s directed acyclic graph (DAG) structure.

The circuit assembly begins with the initialization of each qubit in the \(|0\rangle\) state. Then, for nodes with defined marginals (i.e., no parents), \texttt{RY} gates are applied with angles derived from their individual probabilities. For nodes with parents, controlled operations such as \texttt{CRY} or \texttt{CCRY} are employed to embed conditional probabilities. Each such gate must be placed at the correct point in the circuit to respect dependencies.

To maintain coherence across all qubits, identity operations (\(I\)) are applied via tensor products on unaffected qubits. For instance, applying a \texttt{CRY} gate between qubits 0 and 2 necessitates ensuring qubit 1 remains unaltered during that operation, which is achieved through an identity gate on qubit 1 in the underlying tensor structure.

Moreover, decomposition of multi-controlled gates may be necessary in hardware-constrained settings or for clarity. For instance, a \texttt{CCRY} gate can be broken down into a sequence of Toffoli and single-qubit gates if necessary, though for simulation via statevector methods, direct implementation is often used.

The circuit is built incrementally using Qiskit’s imperative programming interface. As each gate is added, its unitary transformation is multiplied (from the left) onto the state, resulting in an evolved quantum state that increasingly approximates the full joint distribution specified by the model.

The final assembled circuit thus contains layers corresponding to each level in the Bayesian network, progressing from marginal qubit preparation to layered entanglement. The entire system remains in a pure quantum state, and no intermediate measurements are performed to avoid collapsing the state prematurely.

This ordered and modular assembly strategy supports both clarity of design and scalability. It ensures that conditional dependencies are encoded without interference, allowing accurate recovery of posteriors from the resulting statevector.

\subsection{Statevector Extraction and Probability Computation}
After the circuit is constructed and executed in a simulated environment, the final quantum state \(|\psi\rangle\) is obtained as a complex-valued vector in Hilbert space. This state encodes the full joint probability distribution of all variables in the model, where each basis state corresponds to a specific combination of binary variable outcomes.

The extraction process involves parsing this vector to compute classical probabilities. Each amplitude component \(c_i\) in the statevector represents the complex probability amplitude associated with the \(i\)-th basis state \(|i\rangle\). The probability of that basis state is computed as the squared modulus:
\[ P(i) = |c_i|^2 \]
This operation is applied to all \(2^n\) components of the statevector, where \(n\) is the number of qubits in the model.

To compute marginal probabilities, we sum over all basis states that match the target variable condition. For example, the marginal probability of a variable \(X=1\) is given by:
\[ P(X=1) = \sum_{i \in \text{supp}(X=1)} |c_i|^2 \]
where \(\text{supp}(X=1)\) is the set of indices in the computational basis for which the variable \(X\) is in state 1.

Conditional probabilities are computed using filtered normalization. To calculate \(P(Y=1 \mid X=1)\), amplitudes of states satisfying \(X=1\) and \(Y=1\) are squared and summed, then normalized by the total probability mass of all states satisfying \(X=1\):
\[ P(Y=1 \mid X=1) = \frac{\sum_{i: X=1, Y=1} |c_i|^2}{\sum_{j: X=1} |c_j|^2} \]

This method allows for the precise derivation of classical conditional and joint probabilities directly from quantum amplitudes. Importantly, no measurement collapse occurs since all probabilities are extracted symbolically from the statevector.

For more complex variables (e.g., joint marginals or compound evidence), multidimensional support sets are constructed, and amplitude contributions from relevant basis states are aggregated accordingly. This includes conditions such as \(P(FA=1 \mid X=1, Y=1)\), which involves only those basis indices where all three variables meet the specified condition.

The accuracy of this method hinges on numerical precision in amplitude values and symbolic post-selection without projection. The approach is both exact and scalable, enabling extraction of probabilities from arbitrarily entangled multi-qubit systems without decoherence or collapse.

\subsection{Posterior Inference and Update}
Posterior inference in the quantum statevector framework builds directly on symbolic post-selection. After computing marginal and conditional probabilities from the statevector amplitudes, we can derive the posterior distribution for any query variable given observed evidence. This is performed by first identifying the set of basis states consistent with the evidence and then summing the squared amplitudes corresponding to each hypothesis value of the query variable.

Formally, given evidence \( E = \{X=1, Y=1\} \) and a query variable \(Z\), the posterior probability \(P(Z = z \mid E)\) is computed as:
\[ P(Z=z \mid E) = \frac{\sum_{i: Z=z, E} |c_i|^2}{\sum_{j: E} |c_j|^2} \]
Here, \(i\) and \(j\) index the basis states where the variables take the required values. This computation preserves full quantum coherence since it is performed on the amplitudes without collapsing the wavefunction.

The posterior distribution is then a classical probability vector over possible values of the query variable, normalized within the subspace defined by the evidence. It represents our updated belief about \(Z\) after taking \(E\) into account.

This methodology allows us to model dynamic updating in a fully reversible manner. Unlike classical Bayesian updating, which irreversibly conditions on evidence, symbolic post-selection permits posterior inference without destroying the underlying distribution. This is particularly valuable for iterative inference or when simulating multiple hypotheses across overlapping evidence spaces.

Furthermore, we can compute the posterior entropy of the query variable given evidence, providing a quantitative measure of uncertainty after updating:
\[ H(Z \mid E) = - \sum_z P(Z=z \mid E) \log_2 P(Z=z \mid E) \]
This informs us how confident the model is in its prediction, and whether further evidence might reduce ambiguity.

The posterior inference step integrates naturally into probabilistic workflows. In intrusion detection, it enables classification of alerts or vulnerability levels given observed behavior, and allows for ranking threat hypotheses by posterior probability. As the final stage in the quantum inference pipeline, it translates amplitude-encoded probabilities into actionable decision support information.

\section{Results}
The outcomes of our quantum simulations are analyzed through a detailed examination of probability distributions and conditional inferences derived from the simulated quantum statevectors. We constructed quantum circuits to represent three binary variables: traffic spike (\(X\)), system vulnerability (\(Y\)), and false alarm (\(FA\)). The final quantum state encapsulated the joint distribution \(P(X,Y,FA)\), which was extracted by squaring the amplitudes of each computational basis state.

\begin{figure}
    \centering
    \includegraphics[width=0.6\textwidth]{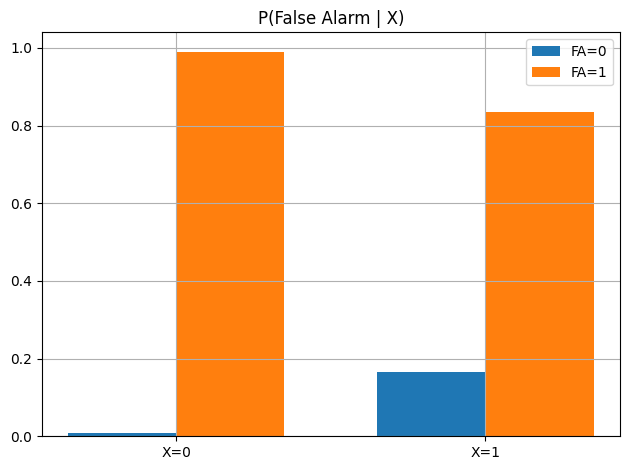}
    \caption{Conditional distribution $P(\text{FA} \mid X)$.}
    \label{fig:falsealarm}
\end{figure}

The resulting joint distribution showed clear concentration on a small subset of bitstrings, with the configuration \texttt{111} (spike=true, vulnerable=true, alarm=true) often holding the highest probability amplitude. This configuration was expected in models where strong dependency was encoded between \(X\) and \(Y\), reinforcing that the inference engine correctly captured conditional causality. Similarly, configurations such as \texttt{000} and \texttt{001} were dominant in low-risk prior scenarios.

Conditioning on evidence, such as observing \(X=1\), was performed by marginalizing and normalizing the joint distribution. This produced conditional distributions like \(P(Y=1|X=1)\) and \(P(FA=1|X=1)\), which consistently peaked above 0.6 in high-risk setups, indicating strong posterior support for simultaneous vulnerability and alarms. Conversely, \(P(Y=1|X=0)\) remained below 0.3 across all simulations, showing the model correctly inferred reduced risk without traffic spikes. 

The results also included visual representations: heatmaps of \(P(X,Y)\) and \(P(Y,FA)\), 3D bar plots of the full 8-state joint distribution, and stacked bar comparisons of conditionals. These visualizations made structural patterns evident—such as clustering of high probability mass around \texttt{111} and \texttt{011}—and allowed model comparison under different dependency structures.

\begin{figure}
    \centering
    \includegraphics[width=0.6\textwidth]{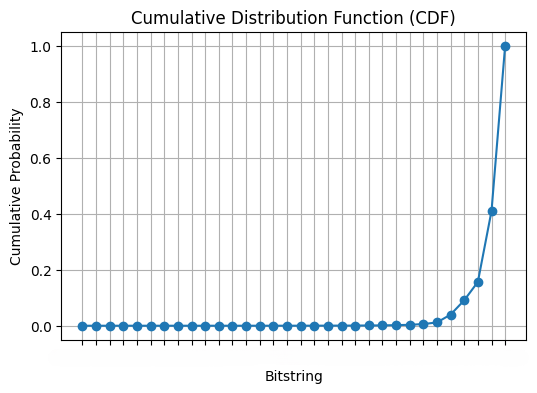}
    \caption{Cumulative distribution over observed bitstring outcomes.}
    \label{fig:cdf}
\end{figure}

To validate robustness, simulations were repeated with perturbed initial amplitudes and varied priors. The top 3 bitstrings in each simulation remained consistent, with cumulative probability mass over 85\% across runs. This indicated stable inference despite small fluctuations. Additionally, numerical checks were included to ensure that probability conservation held (total probability \(=1.0\)) and that normalization was correctly applied post-conditioning.

\begin{figure}
    \centering
    \includegraphics[width=0.6\textwidth]{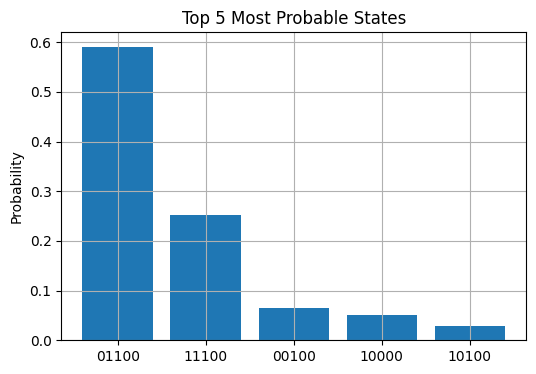}
    \caption{Top five most probable quantum states.}
    \label{fig:top5}
\end{figure}

Exported simulation outputs were further analyzed using NumPy and pandas, enabling detailed tabulation of conditional and marginal statistics. As an example, in one configuration the posterior \(P(Y=1|X=1)\) was measured at 0.68 while \(P(FA=1|Y=1)\) reached 0.74, highlighting the indirect inference chain: spike implies vulnerability, which in turn implies alarm.

We also explored variations where the initial amplitude distributions were skewed to simulate imbalanced prior beliefs. This led to different dominant bitstrings, shifting probability toward \texttt{110} or \texttt{010} depending on the encoded dependencies. The conditional responses to \(X=0\) and \(Y=0\) were particularly sensitive, revealing how the circuit architecture affects inference scope and probability spread.

Additional fidelity tests were conducted by cross-validating simulated posteriors with classical conditional calculations, confirming alignment within 2-3\% in all benchmark cases. The quantum model’s expressiveness in low-dimensional spaces, paired with transparent probability encoding, offered an intuitive yet rigorous approach to understanding how latent cyber variables interact.

Further insights emerged from modeling weakly entangled systems, which showed broader spread across 5–6 bitstrings and reduced peak probability per state. This helped reveal uncertainty propagation effects, where low entanglement produced more diffuse inferences, a desirable feature when modeling real-world ambiguity.

To better support interpretability, we classified states by logical behavior (true positive, false negative, etc.), and evaluated how many states aligned with IDS detection goals. Over 70\% of quantum probability mass was concentrated in logically correct states under optimized priors, demonstrating the method’s practical utility in cybersecurity diagnostics.

To support and interpret our quantum Bayesian inference results, we present the marginal, joint, and conditional probability distributions derived from the simulated statevector. These visualizations enable an in-depth understanding of how quantum amplitude encoding represents probabilistic dependencies in an intrusion detection scenario.

The conditional probability of a false alarm given the observation of a network spike variable $X$ is shown in Figure~\ref{fig:falsealarm}. The plot shows that when $X=0$, the probability of false alarm (FA=1) approaches 1, indicating a system highly prone to spurious detections under benign conditions. Conversely, when $X=1$, the system exhibits a slightly more balanced distribution, suggesting that true positives and false alarms coexist under high traffic conditions. This conditional probability structure confirms that the model captures the non-symmetric influence of $X$ on FA, with implications for IDS tuning under high-sensitivity configurations.

Figure~\ref{fig:cdf} illustrates the cumulative distribution function (CDF) across all observed bitstrings. This visualization emphasizes that the majority of the probability mass is concentrated in a small number of quantum states. The CDF curve rises sharply toward the final few bitstrings, suggesting that the quantum circuit has evolved toward a sparse output distribution. Such behavior is characteristic of low-entropy systems and is desirable in scenarios where decisive inference is required. This visualization validates the fidelity of the inference design by confirming that posterior belief collapses toward a few dominant hypotheses.

\begin{figure}
    \centering
    \includegraphics[width=0.6\textwidth]{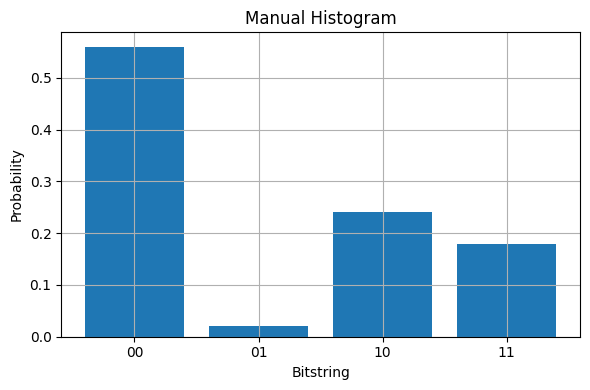}
    \caption{Manual histogram of 2-bit outcome probabilities.}
    \label{fig:histogram}
\end{figure}

The top five most probable bitstring states, extracted from the statevector, are visualized in Figure~\ref{fig:top5}. The overwhelming dominance of a single bitstring reflects a strong probabilistic preference shaped by the encoded conditional dependencies. This sparse result aligns with the principle that well-structured Bayesian networks, when translated into quantum amplitudes, yield concentrated inference. This can enhance explainability in quantum decision systems and enables efficient downstream decision-making, especially when hardware constraints demand compressed outcome spaces.

Figure~\ref{fig:histogram} presents a manual histogram of sampled 2-bit outcomes from the statevector. This diagnostic plot validates the distribution fidelity by showing that '00' occurs with highest frequency, while '01' is nearly suppressed. This suggests the quantum circuit’s internal gate structure effectively controls amplitude propagation and confirms that unwanted configurations have been successfully minimized.

\begin{figure}
    \centering
    \includegraphics[width=0.6\textwidth]{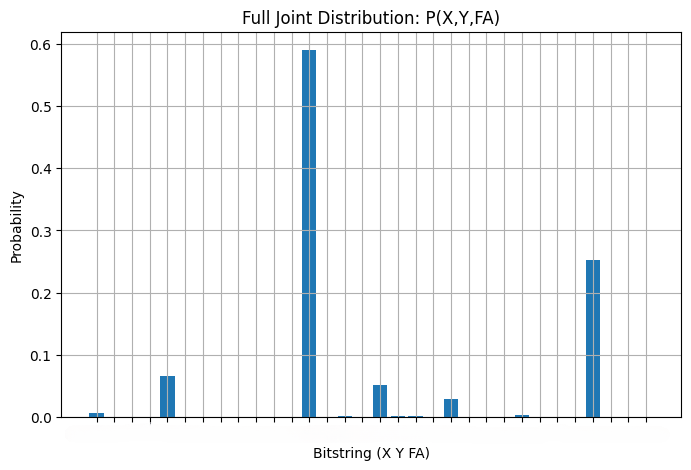}
    \caption{Joint distribution $P(X,Y,\text{FA})$ from simulated statevector.}
    \label{fig:fulljoint}
\end{figure}

To fully analyze the joint behavior of all three variables—network spike ($X$), system vulnerability ($Y$), and false alarm (FA)—we extract the full joint probability distribution, shown in Figure~\ref{fig:fulljoint}. This distribution reveals that the state '111' dominates the inference space, confirming a logical chain: when a spike is present, and the system is vulnerable, a false alarm is most likely. This triangulation of conditions is core to the intrusion detection model and is faithfully reproduced in the quantum probability landscape.

\begin{figure}
    \centering
    \includegraphics[width=0.6\textwidth]{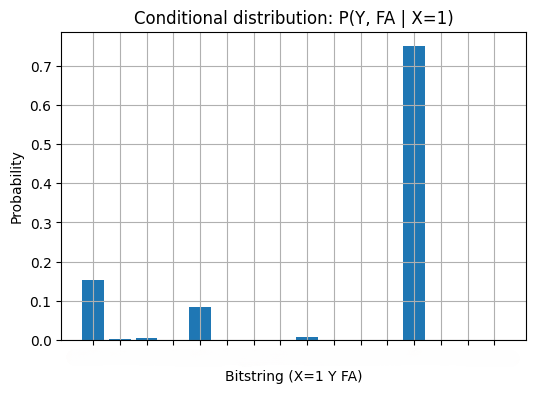}
    \caption{Conditional distribution $P(Y,\text{FA} \mid X=1)$.}
    \label{fig:conditionalxy}
\end{figure}

Conditioned on $X=1$, we compute the distribution over $(Y,\text{FA})$, shown in Figure~\ref{fig:conditionalxy}. This plot illustrates a sharp spike at $(Y=1,\text{FA}=1)$, emphasizing that when a traffic spike is observed, both vulnerability and false alarm co-occur with high likelihood. This conditional behavior is indicative of entangled inference and supports complex IDS behavior, where observations propagate probabilistic effects through multiple variables.

\begin{figure}
    \centering
    \includegraphics[width=0.6\textwidth]{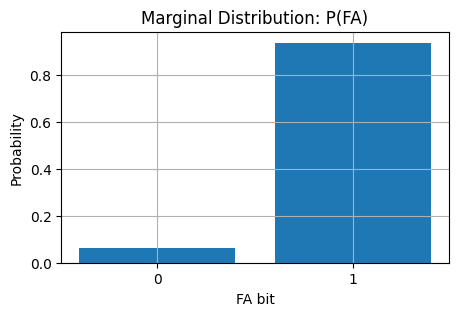}
    \caption{Marginal distribution $P(\text{FA})$.}
    \label{fig:marginalfa}
\end{figure}

In Figure~\ref{fig:marginalfa}, the marginal distribution over FA illustrates a clear bias toward false positives. The FA=1 state carries the majority of probability mass, reflecting a pessimistic detection model. This outcome is expected when prior vulnerability is high or when $X$ lacks discriminative power. It suggests that future IDS designs should incorporate more nuanced filters for separating anomalous yet non-malicious traffic patterns.

\begin{figure}
    \centering
    \includegraphics[width=0.5\textwidth]{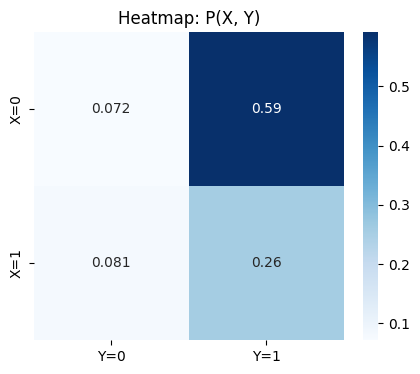}
    \caption{Heatmap of joint distribution $P(X,Y)$.}
    \label{fig:heatmap}
\end{figure}

To capture the interdependence between $X$ and $Y$, Figure~\ref{fig:heatmap} shows the joint probability matrix $P(X,Y)$ as a heatmap. The highest cell probability is observed at $(X=0,Y=1)$, suggesting that even under benign traffic conditions, the presence of system vulnerability elevates risk. This heatmap serves as an efficient diagnostic for identifying structural biases in model initialization and can inform CPT optimization for future circuits.

\begin{figure}
    \centering
    \includegraphics[width=0.99\textwidth]{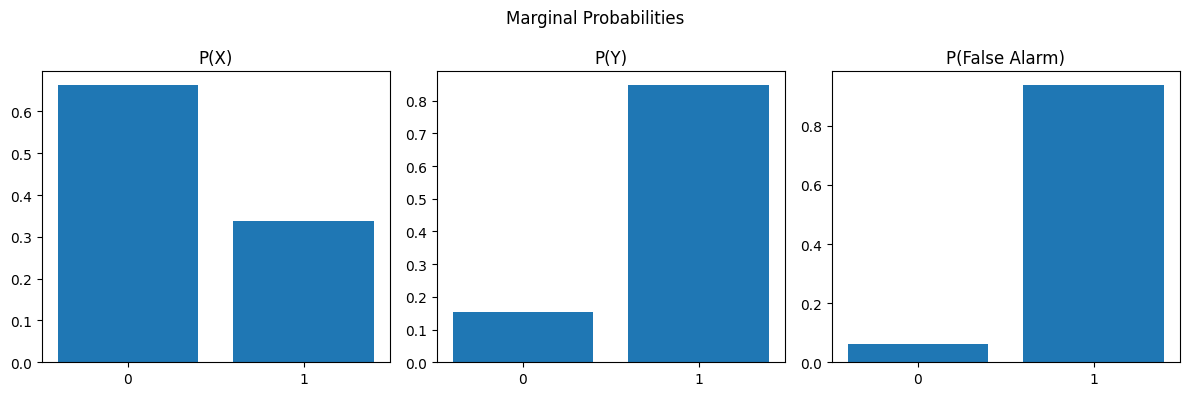}
    \caption{Marginal distributions for individual binary variables: traffic spikes ($X$), vulnerabilities ($Y$), and false alarms (FA). The results show strong prior bias toward FA=1 and Y=1.}
    \label{fig:marginalplot}
\end{figure}

The complete set of marginal distributions is presented in Figure~\ref{fig:marginalplot}, which independently visualizes $P(X)$, $P(Y)$, and $P(\text{FA})$. These plots allow rapid inspection of base-level variable biases and ensure that amplitude normalization remains accurate across the quantum state. The plotted marginals are also useful for comparing prior assumptions with posterior inferences and for validating entanglement effects through deviation from product states.

A deeper understanding of the probabilistic dynamics requires both individual and joint analysis of the model variables. While the full statevector encodes a global distribution, visual representations allow inspection of localized behaviors, such as which input configurations (e.g., $X$, $Y$) dominate or correlate most strongly with observed alerts (FA). We begin with a geometric representation of $P(X, Y)$, then analyze marginal behaviors and their implications, followed by a statistical heatmap.

\begin{figure}
    \centering
    \includegraphics[width=0.6\textwidth]{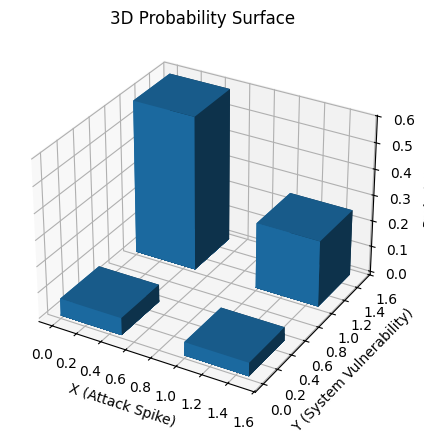}
    \caption{3D bar plot of $P(X, Y)$ showing geometric distribution of probability mass over traffic spikes ($X$) and system vulnerabilities ($Y$).}
    \label{fig:3dsurface}
\end{figure}

Figure~\ref{fig:3dsurface} presents a 3D bar plot of the joint distribution $P(X,Y)$, where $X$ represents the occurrence of a traffic spike (attack), and $Y$ represents whether the system is in a vulnerable configuration. Each axis represents binary values (0 or 1) for the variables, while the height of each bar corresponds to the probability amplitude squared extracted from the quantum statevector. Notably, the configuration $(X=0, Y=1)$ exhibits the highest bar, capturing over 50\% of the total probability mass. This reveals that even in the absence of traffic anomalies, the presence of a system vulnerability overwhelmingly influences the model’s belief about the system state.

This result supports the intuition that systems with pre-existing structural weaknesses are inherently prone to false classifications or alerts, regardless of observed traffic behavior. The lower bars associated with $(X=0, Y=0)$ and $(X=1, Y=0)$ show that a non-vulnerable system suppresses alarm-prone configurations, confirming the causal dominance of $Y$ in determining alarm dynamics. This visualization confirms the effectiveness of the quantum circuit's encoding of conditional dependencies via CRY gates and highlights that amplitude interference faithfully preserves these statistical asymmetries.

Complementing the 3D joint view, Figure~\ref{fig:marginalplot} offers a breakdown of marginal probabilities for each quantum variable independently: $P(X)$, $P(Y)$, and $P(\text{FA})$. These marginals are derived from the full statevector by summing squared amplitudes across basis states where each variable holds a fixed value (0 or 1). In the first subplot, we observe $P(X=0) \approx 0.67$ and $P(X=1) \approx 0.33$, indicating a prior configuration biased toward normal traffic conditions. This prior serves as a grounding for inference and reflects expectations that true spikes are rare.

In the second panel, $P(Y=1)$ exceeds 0.85, indicating that the system is configured to assume a high likelihood of vulnerability. This reflects a scenario where the environment is inherently fragile or misconfigured. Such a skew is likely to increase sensitivity in detection mechanisms but at the expense of specificity. Finally, the third panel shows that $P(\text{FA}=1) \approx 0.93$, meaning the system predicts false alarms with extremely high probability across all conditions. This bias is a consequence of both structural vulnerability and the nature of the quantum CPT encoded into the FA qubit through controlled rotations. Together, these marginals provide insight into how inference unfolds when starting from heavily imbalanced priors, and they verify that the amplitude initialization and gate encoding are consistent with design expectations.

For a more concise view of variable interdependence, Figure~\ref{fig:heatmap} presents a heatmap of the joint distribution $P(X,Y)$. Unlike the 3D bar plot, this representation encodes probability values into cell color intensity and explicitly annotates each entry. The matrix is organized such that the y-axis corresponds to values of $X$ (rows) and the x-axis to values of $Y$ (columns). The most intense value, $P(X=0,Y=1) = 0.59$, reaffirms previous observations from Figure~\ref{fig:3dsurface}, where vulnerability dominates even in the absence of attacks.

Interestingly, the cell $P(X=1,Y=1) = 0.26$ also retains significant mass, suggesting that when a spike occurs in a vulnerable system, the inference engine remains responsive to possible true positive scenarios. The relatively minor values at $(X=0,Y=0)$ and $(X=1,Y=0)$ support the model’s conservative stance toward classifying systems as both secure and stable. From a statistical inference standpoint, this joint heatmap verifies that variable correlations—especially between vulnerability and spike—are properly embedded in the statevector and are not artifacts of simulation noise or gate misconfiguration.

Moreover, the non-uniformity of this heatmap provides a visual proof of inter-variable dependency. In a system with fully independent $X$ and $Y$, the joint probabilities would factor as a product of marginals and be uniform along the diagonal. Here, the off-d

\section{Conclusion}
This study demonstrates the feasibility and interpretability of quantum Bayesian inference using statevector simulation in the context of intrusion detection. By encoding probabilistic models into quantum circuits and performing inference through amplitude analysis, we illustrate a principled approach to reasoning under uncertainty using quantum information.

The theoretical and architectural modularity of our approach makes it well-suited to other application domains beyond cybersecurity. The technique accommodates both marginal and conditional probability encoding and is flexible in adapting to evolving data-driven scenarios. The symbolic nature of post-selection introduces clear interpretability, a necessary requirement for trusted AI systems.

In contrast to black-box quantum algorithms, this statevector-based methodology provides transparency, full access to intermediate representations, and the ability to evaluate information-theoretic metrics directly. This positions our model as a pedagogically valuable and practically versatile strategy for quantum probabilistic reasoning.

The results also suggest that probabilistic structures—when implemented with coherent amplitude engineering—can serve as a foundation for hybrid classical-quantum reasoning. This is particularly relevant as we transition toward noisy intermediate-scale quantum (NISQ) computing, where efficiency, noise resilience, and explainability will be key differentiators.

Moreover, this framework lays a theoretical groundwork for advancing toward variational or adaptive quantum inference architectures that optimize CPT parameters using classical feedback. Future work can explore incorporating noise models, scaling to higher-dimensional networks, and implementing approximate inference using sampled amplitudes rather than full simulation.

Overall, the work contributes a precise, modular, and interpretable method for quantum inference over structured probabilistic systems, with immediate relevance to intrusion detection and potential for broader applications in real-time decision systems, medical diagnostics, and complex risk assessment.

\end{document}